\documentclass[12pt]{article}

\usepackage{epsfig,amstext,amsthm, amsmath, amsfonts, a4, color,amssymb}
\usepackage{latexsym}
\usepackage{natbib}
\usepackage{graphicx}
\usepackage{placeins, bm}

\def\hang{\hangindent\parindent}
\def\rf{\par\noindent\hang}

\newcommand{\E}{\operatorname{E}}

\renewcommand{\mathbf}{\boldsymbol}

\renewcommand{\tilde}[1]{\widetilde{#1}}
\renewcommand{\hat}[1]{\widehat{#1}}

\newcommand{\ba}{{\mathbf{a}}} 

\newcommand{\cc}{{\mathbf{c}}}

\newcommand{\ee}{{\mathbf{e}}}

\newcommand{\II}{{\mathbf{I}}}

\newcommand{\MM}{{\mathbf{M}}}

\newcommand{\XX}{{\mathbf{X}}}
\newcommand{\YY}{{\mathbf{Y}}}

\newcommand{\be}{{\mbox{\boldmath $\beta$}}}

\newcommand{\et}{{\mbox{\boldmath $\eta$}}}

\newcommand{\ps}{{\mbox{\boldmath $\psi$}}}

\newcommand{\vare}{{\mbox{\boldmath $\varepsilon$}}}

\newcommand{\zer}{{\mbox{\boldmath $0$}}}

\begin{document}
\title{Fletcher-Turek Model Averaged Profile Likelihood Confidence Intervals}
\author{Paul Kabaila\\
Department of Mathematics and Statistics,
La Trobe University\\
A.H. Welsh\\
Mathematical Sciences Institute,
The Australian National University\\
Waruni Abeysekera\\
Department of Mathematics and Statistics,
La Trobe University
}

\date{\today}

\maketitle

Running headline: Model Averaged Confidence Intervals

\abstract{
We evaluate the model averaged profile likelihood confidence intervals proposed by Fletcher and Turek (2011) in a simple situation in which there are two linear regression models over which we average.  We obtain exact expressions for the coverage and the scaled expected length of the intervals and use these to compute these quantities in particular situations. We show that the Fletcher-Turek confidence intervals can have coverage well below the nominal coverage and expected length greater than that of the standard confidence interval with coverage equal to the same minimum coverage.  In these situations, the Fletcher-Turek confidence intervals are unfortunately not better than the standard confidence interval used after model selection but ignoring the model selection process.}


\thispagestyle{empty}

\pagenumbering{arabic}
\setcounter{page}{1}

\section{Introduction}

\noindent It is common practice in applied statistics to carry out data-based model selection by, for example, using preliminary hypothesis tests or minimizing a criterion such as the Akaike Information Criterion (AIC) and then to use the selected model to construct confidence intervals as if it had been given to us {\it a priori} as the true model. This procedure can lead to confidence intervals with minimum coverage probabilities far below the nominal coverage probability; see Kabaila (2009) for a review of the literature on this topic.

In recent years, there has been growing interest in using techniques which involve several models to try to incorporate model uncertainty into the inferences.  These techniques, loosely referred to as model-averaging, are used in both the Bayesian and the frequentist literature; see, for example, Buckland et al. (1997), Raftery et al. (1997), Volinsky et al. (1997), Hoeting et al. (1999), Burnham and Anderson (2002) and Claeskens and Hjort (2008).  In this paper, we focus on frequentist model-averaging techniques for constructing confidence intervals.

The earliest frequentist approach to constructing model-averaged confidence
intervals (see Buckland et al, 1997 and Burnham and Anderson, 2002)
was to centre the interval on a model-averaged estimator and determine the width
of the interval by an estimate of the standard deviation of this
estimator.  The distribution theory on which these intervals are based is not (even approximately) correct (Claeskens and Hjort, 2008, p.207) but simulation studies report that these intervals work well in terms of coverage
probability in particular cases (Lukacs et al., 2010; Fletcher and Dillingham, 2011).
A different approach was proposed by Hjort and Claeskens (2003) but this turns out to be essentially the same as the standard confidence interval based on fitting a full model (Kabaila and Leeb, 2006; Wang and Zou, 2013).  More recently, Fletcher
and Turek (2011) and Turek and Fletcher (2012) have proposed averaging confidence interval construction procedures from each of the possible models. Fletcher and Turek (2011) averaged the profile likelihood confidence interval procedure and Turek and Fletcher (2012) averaged the tail areas of the distributions of the estimators from each of the possible models.

Given the practical importance of the problem, it is not surprising that considerable hope has been invested in model averaging as a simple, general method for making valid inferences under model uncertainty.  In this context, it is important to develop a theoretical understanding of the properties of model averaging procedures so that we can put their increasing use on a firm basis.  A good starting point is to explore the properties of procedures in meaningful, tractable scenarios which allow us to evaluate whether they work as expected, to compare different proposals and perhaps to modify and improve current proposals.  We make a start on this by studying the theoretical properties of the Fletcher and Turek (2011) model averaged profile likelihood confidence interval procedure in a simple scenario that is both meaningful and tractable.

We obtain a $1-\alpha$ level profile likelihood confidence interval for a parameter $\theta$ in a model ${\cal M}_j$ by computing the signed-root log-likelihood ratio for $\theta$ under ${\cal M}_j$ and then solving for the lower and upper endpoints of the interval the two equations obtained by equating the normal cumulative distribution function evaluated at the signed-root log likelihood ratio to $1-\alpha/2$ and $\alpha/2$, respectively. When we have models $\{{\cal M}_1,\ldots, {\cal M}_R\}$ for a fixed, finite $R$, the  Fletcher and Turek (2011) model averaged profile likelihood  confidence interval (MPI) for $\theta$, with nominal coverage $1-\alpha$, is obtained by solving for the endpoints a weighted average of the profile likelihood confidence interval endpoint equations for each model.  There are various ways to choose the weights; Fletcher and Turek focus on weights derived by exponentiating the Akaike Information Criterion (AIC) for each model.

The only evaluation of the MPI to date has been by simulation; Fletcher and Turek (2011) showed that the MPI performs well in particular settings.  It is natural to use simulations to evaluate different confidence intervals, but simulation methods have weaknesses for evaluating performance criteria.  First, simulations cover only a limited set of particular settings (particularly, values of the unknown nuisance parameters) and the conclusions apply only to these settings.  They may therefore not consider settings where the coverage is low or the expected length is large.  We can improve the situation by evaluating minimum coverage probabilities and maximum expected lengths to characterise performance over unknown nuisance parameters.  Secondly, the variability in simulation results complicates finding bounds on coverage or expected length, particularly when there are a large number of parameters to vary in the underlying distribution.  We therefore use exact calculations to evaluate the properties of the confidence intervals both in particular settings and uniformly over unknown nuisance parameters.

For simplicity, we consider a scenario with only two possible models, a linear regression model with independent and identically distributed normal errors (${\cal M}_2$) and the same model with a linear constraint on the regression parameters (${\cal M}_1$).  We evaluate the properties of the MPI, with nominal coverage $1-\alpha$, for a parameter of interest $\theta$ that is common to both models.  This scenario is simple but, nonetheless, includes practically important problems. For example, in the comparison of two treatments for a given value of the single covariate in a one-way analysis of covariance, the parameter of interest $\theta$ is the treatment effect for a given value of the covariate and the two models ${\cal M}_2$ and ${\cal M}_1$ are distinguished by whether $\tau$, the difference in the coefficients of the covariate, is unconstrained or constrained to equal zero (so the fitted models have parallel mean functions).  In general, $\theta$ and $\tau$ can be any linearly independent linear functions of the regression parameter and we obtain general results for any given model matrix, so allowing any possible set of nuisance regression parameters.  We focus on two properties, the coverage and the scaled expected length, where the scaling is with respect to the length of the standard confidence interval at the minimum coverage level.  We derive computationally convenient, exact expressions for the coverage probability and the scaled expected length of the MPI for $\theta$, so that we do not need to resort to simulations.

Our results show that there are situations in which the MPI would be expected to work well but has poor coverage, much lower than the nominal coverage, and expected length greater than that of the standard confidence interval with coverage equal to the minimal coverage.  In these situations, the MPI performs worse than standard confidence intervals used after model selection but ignoring the model selection process.  While disappointing result undermines the hope that the MPI could be generally applicable, it reinforces the the need to develop new procedures and highlights the need for careful analysis of new procedures.

We present our theoretical results in Section 2 and illustrate their application to a real data example from a cloud seeding experiment in which the parameter of interest is the effect of cloud seeding in Section 3.  We present the coverage probability and the scaled expected length of the MPI for the parameter of interest and show how to interpret these values.  We conclude with a brief discussion in Section 4.  Theoretical calculations and the proofs of the Theorems are presented in an Appendix.

%
%
%
%

\section{Theoretical details} \label{sec:four}

In this Section, we describe how to compute the profile likelihood confidence interval for $\theta$ and the MPI for $\theta$, and then give exact theoretical expressions for the coverage and the scaled expected length of these intervals. The proofs are left to the Appendix.

The model ${\cal M}_2$ is given by
\[
\YY = \XX \be + \vare,
\]
where $\YY$ is a random $n$-vector of responses, $\XX$ is a known
$n\times p$ model matrix with $p < n$ linearly independent columns, $\be$ is
an unknown $p$-vector parameter  and $\vare \sim \text{N}(\zer, \sigma^2 \II_n)$,
with $\sigma^2$ an unknown positive parameter. Suppose that we are interested in making inference about the parameter $\theta = \ba^{\top} \be$,  where $\ba$ is a specified nonzero $p$-vector.  Suppose also that we define the parameter $\tau = \cc^{\top} \be - t$,  where $\cc$ is a specified nonzero $p$-vector that is linearly independent of $\ba$ and $t$ is a specified number.  The model ${\cal M}_1$ is ${\cal M}_2$ with $\tau=0$.

Let $\hat{\be}$ be the least squares estimator of $\be$ and $\hat{\sigma}^2 = (\YY-\XX\hat{\be})^{\top}(\YY-\XX\hat{\be})/(n-p)$ be the usual unbiased estimator of $\sigma^2$.  Set $\hat{\theta} = \ba^{\top} \hat{\be}$ and $\hat{\tau} = \cc^{\top} \hat{\be} - t$. Define $v_{\theta} = \ba^{\top}(\XX^{\top}\XX)^{-1}\ba$ and $v_{\tau} = \cc^{\top}(\XX^{\top}\XX)^{-1}\cc$.  Then two important quantities are the known correlation $\rho=\ba^{\top}(\XX^{\top}\XX)^{-1}\cc/(v_{\theta}v_{\tau})^{1/2}$ between $\hat{\theta}$ and $\hat{\tau}$ and the unknown parameter $\gamma = \tau \big/\big(\sigma v_{\tau}^{1/2} \big)$.


We adopt the definition of signed-root log-likelihood ratio statistic used by Fletcher and Turek (2011). This is {\sl minus} the usual definition;
which definition we adopt makes no essential difference to the results.
We show in the Appendix that the signed-root log-likelihood ratio statistic for ${\cal M}_2$ is $r_2 \big\{(\hat{\theta}-\theta)/v_{\theta}^{1/2}, \, \hat{\sigma} \big\} = r_2 \big\{(\hat{\theta}-\theta)/(\sigma v_{\theta}^{1/2}), \, \hat{\sigma}/\sigma \big \}$, where
\begin{eqnarray}
r_2(\delta,y) &=& \text{sign}(\delta)\left[n\log \left\{1+ \frac{\delta^2}{(n-p) y^2}\right\}\right]^{1/2}, \label{eq:r2}
\end{eqnarray}
and the signed-root log-likelihood ratio statistic for ${\cal M}_1$ is $r_1 \big\{(\hat{\theta}-\theta)/v_{\theta}^{1/2}, \, \hat{\tau}/v_{\tau}^{1/2}, \, \hat{\sigma} \big \} = r_1 \big\{(\hat{\theta}-\theta)/(\sigma v_{\theta}^{1/2}), \, \hat{\tau}/(\sigma v_{\tau}^{1/2}), \, \hat{\sigma}/\sigma \big\}$, where
\begin{eqnarray}
r_1(\delta, x, y)
&=& \text{sign}(\delta-\rho x)\left(n\log\left[1+\frac{(\delta  - \rho x)^2}{(1-\rho^2)\{x^2 + (n-p) y^2 \}}\right]\right)^{1/2}. \label{eq:r1}
\end{eqnarray}

We can derive a profile likelihood confidence interval for $\theta$ from the models ${\cal M}_2$ and ${\cal M}_1$ separately or from a weighted average of the profile likelihood confidence endpoint equations for the
models ${\cal M}_2$ and ${\cal M}_1$.  Let $\Phi$ denote the standard normal cumulative distribution function.  Then the profile likelihood confidence interval for $\theta$ from model ${\cal M}_2$, with nominal coverage $1-\alpha$, is $\big[\hat{\theta}_{2l},\, \hat{\theta}_{2u}\big]$, where $\hat{\theta}_{2l} < \hat{\theta}$ solves in $\theta$ the equation
\begin{eqnarray*}
\Phi \left[r_2 \left\{(\hat{\theta}-\theta)/v_{\theta}^{1/2},\, \hat{\sigma} \right\} \right] &=& 1-\alpha/2
\end{eqnarray*}
and $\hat{\theta}_{2u} > \hat{\theta}$ solves in $\theta$ the equation
\begin{eqnarray*}
\Phi \left[r_2 \left\{(\hat{\theta}-\theta)/v_{\theta}^{1/2},\, \hat{\sigma} \right\} \right] &=& \alpha/2.
\end{eqnarray*}


The MPI is obtained by averaging the equations defining the profile likelihood intervals under the models $\mathcal{M}_2$ and $\mathcal{M}_1$.  Fletcher and Turek (2011) focus on the Akaike weights which, for the models ${\cal M}_1$ and ${\cal M}_2$, are $w_1$ and $1-w_1$ respectively, where
$w_1=w_1\big(\hat{\tau}/v_{\tau}^{1/2}, \hat{\sigma} \big) = w_1\big(\hat{\tau}/(\sigma v_{\tau}^{1/2}), \, \hat{\sigma}/\sigma \big)$, with
\begin{equation}
w_1(x,y)= \frac{1}{1 + \Big\{1+\frac{x^2}{(n-p)y^2}\Big\}^{n/2} \exp(-1)}. \label{eq:weight}
\end{equation}
We can consider other weights, including weights obtained by replacing AIC by other model selection criteria.  We follow Fletcher and Turek for now and discuss the effect of changing the weights below.  For each $x \in R$ and $y > 0$, define
\begin{eqnarray}
h(\delta,x,y) &=& w_1(x,y) \, \Phi\{r_1(\delta,x,y)\} + \{1-w_1(x,y)\} \, \Phi\{r_2(\delta,y)\} \label{eq:h}
\end{eqnarray}
with $r_2$, $r_1$ and $w_1$ defined by (\ref{eq:r2})--(\ref{eq:weight}).  The MPI,
with nominal coverage $1-\alpha$, is $ \left[\hat{\theta}_l, \,  \hat{\theta}_u \right]$, where
$\hat{\theta}_{l} < \hat{\theta}$ and $\hat{\theta}_{u} > \hat{\theta}$ satisfy
\[
h \left\{(\hat{\theta}-\hat{\theta}_l)/v_{\theta}^{1/2}, \hat{\tau}/v_{\tau}^{1/2}, \hat{\sigma} \right\} = 1-\alpha/2 \ \ \ \mbox{ and } \ \ \ h \left\{(\hat{\theta}-\hat{\theta}_u)/v_{\theta}^{1/2}, \hat{\tau}/v_{\tau}^{1/2}, \hat{\sigma} \right\} = \alpha/2
\]
respectively.

We are interested in the coverage and expected length properties of the MPI of Fletcher and Turek (2011).  For each $x \in R$ and $y > 0$, define $\delta_{u}(x,y)$ to be the solution in $\delta$ of  the equation $h(\delta, x,y) = u$, where $h$ is defined by (\ref{eq:h}).
Theorems 1--3 below on the properties of the MPI are used to construct Figures 1--7 in the next Section; the proofs are given in the Appendix.

\smallskip

\noindent \textbf {Theorem 1}  \textit{The coverage probability of the MPI (averaged over ${\cal M}_1$ and ${\cal M}_2$), with nominal coverage $1-\alpha$, is
\begin{eqnarray*}
P \left(\hat{\theta}_l \le \theta \le \hat{\theta}_u \right) &=&
\int_{0}^{\infty} \int_{-\infty}^{\infty} \Bigg[ \Phi\Bigg\{\frac{\delta_{1-\alpha/2}(x,y) - \rho (x-\gamma)}{(1-\rho^2)^{1/2}} \Bigg\} \\
&& - \Phi\Bigg\{\frac{\delta_{\alpha/2}(x,y) - \rho (x-\gamma)}{(1-\rho^2)^{1/2}} \Bigg\}\Bigg] \phi(x-\gamma) \, f_{n-p}(y) \, dx \, dy,
\end{eqnarray*}
where $\phi$ is the probability density function of the standard normal distribution and $f_{\nu}(y)$ is the probability density function of
$(Q/\nu)^{1/2}$, where $Q$ has a $\chi^2_{\nu}$ distribution.}

\smallskip

\noindent Theorem 1 shows that the coverage of the MPI (averaged over ${\cal M}_1$ and ${\cal M}_2$) is a function of the nominal coverage $1-\alpha$, the residual degrees of freedom $n-p = n(1-p/n)$, the correlation $\rho$ between $\hat{\theta}$ and $\hat{\tau}$, and the unknown parameter
$\gamma=\tau \big/ \big(\sigma v_{\tau}^{1/2}\big)$.  The only unknown quantity is $\gamma$.  We use the minimum coverage over $\gamma$ to describe the worst case results without having to specify particular values for $\gamma$.

We can relate the coverage probability of the MPI to that of the profile likelihood confidence interval under ${\cal M}_2$ and obtain a very useful upper bound to the minimum coverage probability.

\smallskip

\noindent \textbf {Corollary 1}  \textit{The coverage probability of the MPI (averaged over ${\cal M}_1$ and ${\cal M}_2$), with nominal coverage $1-\alpha$, converges to the coverage probability of the profile likelihood interval under
${\cal M}_2$, with nominal coverage $1-\alpha$, as $\gamma \rightarrow \infty$. That is,
\begin{equation*}
\label{corr1_result}
P\left(\hat{\theta}_l \le \theta \le \hat{\theta}_u \right)
\rightarrow
P\left(\hat{\theta}_{2l} \le \theta \le \hat{\theta}_{2u} \right)
\ \ \ \text{as} \ \ \ \gamma \rightarrow \infty.
\end{equation*}
An immediate consequence is that
%
%
\begin{equation*}
\inf_{\gamma} P\left(\hat{\theta}_l \le \theta \le \hat{\theta}_u \right)
\le
P\left(\hat{\theta}_{2l} \le \theta \le \hat{\theta}_{2u} \right) = 2 \, G_{n-p}\left [ (n-p)^{1/2} \left\{\exp \left(\frac{z_{1-\alpha/2}^2}{n} \right)-1\right\}^{1/2} \, \right] - 1,
\end{equation*}
where $G_{n-p}$ denotes the distribution function of the Student t distribution with $n-p$ degrees of freedom
and $z_{1-\alpha/2} = \Phi^{-1}(1-\alpha/2)$.
}


%
%
%

\smallskip

Corollary 1 shows that the minimum coverage probability of the MPI cannot be better than the coverage probability of the profile likelihood interval under $\mathcal{M}_2$.  (Of course, it could be worse.)  In effect, if the profile likelihood interval under $\mathcal{M}_2$ has poor coverage, this will be inherited by the MPI.
Perhaps surprisingly, the coverage of the profile likelihood interval can be well below the nominal level $1-\alpha$.  To see this
note that for fixed $p/n=r$, the upper bound on the coverage probability is
\begin{eqnarray*}
\lefteqn{2 \, G_{n(1-r)}\left [ n^{1/2}(1-r)^{1/2} \left\{\exp \left(\frac{z_{1-\alpha/2}^2}{n} \right)-1\right\}^{1/2} \, \right] - 1 }\\
&=& 2 \, G_{n(1-r)}\left [ (1-r)^{1/2} \left\{z_{1-\alpha/2}^2 + O(n^{-1})\right\}^{1/2} \, \right] - 1\\
&\rightarrow & 2 \Phi\left\{ (1-r)^{1/2} z_{1-\alpha/2}\right\} - 1 \, , \ \ \ \text{as} \ \ n \rightarrow \infty.
\end{eqnarray*}
Thus the coverage probability of the profile likelihood confidence interval under ${\cal M}_2$ decreases as $p/n=r$ increases and is substantially less than the nominal coverage $1-\alpha$ unless $p/n$ is small.  Corollary 1 shows that the MPI will also have poor coverage properties unless $p/n$ is small.

For the expected length of the MPI, we obtain the following result.

\smallskip

\noindent \textbf {Theorem 2}  \textit{The expected length of the MPI (averaged over ${\cal M}_1$ and ${\cal M}_2$), with nominal level $1-\alpha$, is
\begin{eqnarray*}
\E \left(\hat{\theta}_u- \hat{\theta}_l \right)
&=& \sigma \, v_{\theta}^{1/2} \int_{0}^{\infty} \int_{-\infty}^{\infty}
\left \{\delta_{1-\alpha/2}(x,y) - \delta_{\alpha/2}(x,y) \right \} \phi(x-\gamma) \, f_{n-p}(y) \, dx \, dy,
\end{eqnarray*}
where $\phi$ is the probability density function of the standard normal distribution and $f_{\nu}(y)$ is the probability density function of
$(Q/\nu)^{1/2}$, where $Q$ has a $\chi^2_{\nu}$ distribution.}

\smallskip

\noindent Let $c_{\text{min}}$ denote the minimum coverage probability of the MPI (averaged over ${\cal M}_1$ and ${\cal M}_2$), with nominal coverage $1-\alpha$.
The scaled expected length of this confidence interval is therefore
\begin{align*}
&\frac{\E \big(\hat{\theta}_{u} - \hat{\theta}_{l} \big )}{2 \, G_{n-p}^{-1}((c_{\text{min}}+1)/2) \, \E(\hat{\sigma}) \, v_{\theta}^{1/2}} \\
&= \frac{\E(\hat{\theta}_{u} - \hat{\theta}_{l})}{2 \, G_{n-p}^{-1}((c_{\text{min}}+1)/2) \, \sigma \, v_{\theta}^{1/2}
\int_{0}^{\infty} y f_{n-p}(y) dy}\\
 &= \frac{\int_{0}^{\infty} \int_{-\infty}^{\infty} \{\delta_{1-\alpha/2}(x,y) - \delta_{\alpha/2}(x,y)\} \phi(x-\gamma) \, f_{n-p}(y) \,dx \, dy}
 {2 \, G_{n-p}^{-1}((c_{\text{min}}+1)/2) \, \int_{0}^{\infty} yf_{n-p}(y) dy}.
\end{align*}
The integral in the denominator has the analytic expression $2^{1/2} \, \Gamma\{(n-p+1)/2\}\big/ \big[(n-p)^{1/2} \, \Gamma\{(n-p)/2\} \big]$.  As with the coverage, the only unknown quantity in this expression is  $\gamma$, so we study the maximum
scaled expected length over $\gamma$.

The range of calculations needed to evaluate the coverage probability and the scaled expected length of the MPI are reduced by the following result that shows that, because of symmetry, we need only consider $\gamma \ge 0$ and $\rho \ge 0$.

\noindent \textbf {Theorem 3}  \textit{The coverage probability and the scaled expected length of the MPI (averaged over ${\cal M}_1$ and ${\cal M}_2$) are both even functions of $\gamma$ for fixed $\rho$ and even functions of $\rho$ for fixed $\gamma$.}

As we noted above, we can replace AIC by other model selection criteria in the weights.  A convenient way to do this is to replace the penalty $2 \times (\text{\# regression parameters})$ in AIC by $d\times (\text{\# regression parameters})$, where, for some $0 \le u \le 1$,
\[
d  = n \log\left[1 + \frac{\big\{G_{n-p}^{-1}(1-u/2) \big\}^2}{n-p}\right] \rightarrow z_{1-u/2}^2,\,\,\,\mbox{ as } n \rightarrow \infty.  
\]
Here $u$ is the significance level of the equivalent test for the significance of an additional parameter; see the Appendix for more details.  In this case, the $\exp(-1)$ term in the Akaike weights is replaced by $\exp(-d/2)$.  Using the asymptotic approximation to $d$, we find that AIC corresponds to $u=0.157$; more extreme examples (lower significance level) are the usual $u=0.05$ level which gives $\exp(-3.84/2) \approx \exp(-1.92)$ and the Bayesian Information Criterion (BIC) $u=2[1-\Phi\{(\log n)^{1/2}\}]$ which gives $\exp\{-\log(n)/2\}=n^{-1/2}$; less extreme examples (higher significance level) such as $u=0.5$ which gives $\exp(-0.45/2)\approx \exp(-0.227)$ can also be considered.  

We explored the effect of  changing $d$, hoping in particular that values of $d < 2$ might improve the performance of the MPI when $p/n$ is not small, but this is not the case and changing $d$ has very little effect.  Theoretical support for this conclusion is provided by noting that Corollary 1 holds for any fixed value of $d \ge 0$ so, irrespective of the fixed value of $d \ge 0$, the minimum coverage probability of the MPI, with nominal coverage $1-\alpha$,  cannot exceed the coverage probability of the profile likelihood confidence interval, with nominal coverage $1-\alpha$, using  ${\cal M}_2$. Our conclusion is that changing $d$ does not change the ``$p/n$ not small'' problem.  In the boundary case, $d=0$, we have no penalty on the number of regression parameters so we might expect MPI to always use the model ${\cal M}_2$. However, the weight reduces to
\begin{equation}
w_1 = \frac{1}{1 + \left(1 + \frac{\hat{\tau}^2}{(n-p) \hat{\sigma}^2 v_{\tau}} \right)^{n/2}}
\end{equation}
in this case and we still average over the two models. Similarly, for each fixed $d \ge 0$, we do not recover the profile likelihood confidence interval as $\gamma \rightarrow 0$ or even $\hat{\gamma} \rightarrow 0$, but continue to average over the two models.

\section{Cloud seeding example}\label{sec:example}

In this Section, we illustrate how we can use our results on the properties of the MPI in the context of a real data example from a cloud seeding experiment.  The data are presented and analysed by Biondini, Simpson and Woodley (1997), Miller (2002, Section 3.12) and Kabaila (2005). Following Kabaila (2005), we compare the effect of seeding (TRT=1) against the random control (TRT=2) treatment in the moving echo motion category (CAT=1) subgroup of the data. The response variable is the floating target rainfall volume ($m^3 \times 10^7$) and the sample size is $n=33$.  In addition to the treatment indicator, there are five other predictor variables:  coverage (percent) which measures the cloud cover in the target area; seedability  (km); prewetness ($m^3 \times 10^7$) which measures the rainfall in the target area in the hour before treatment; earliness  (hrs) which measures the number of hours in the morning in which there were clouds in the target area; and the average speed of echo motion (knots).  The models considered by Miller (2002, Section 3.12) and Kabaila (2005) included the intercept, treatment indicator, the main effects, squared effects and the interactions between the five predictor variables so that $p$, the dimension of the regression parameter vector, is 22.    All these additional variables can be included in the model or not; variable selection has been carried out by Miller (2002, Section 3.12) and Kabaila (2005) for many variables in this study.  For illustration, we consider model averaging over the full model ($p=22$) and the submodel excluding the squared seedability term $s^2$ whose coefficient we denote by $\tau$.   The goal is to construct a $95\%$ confidence interval for $\theta$, the expected response when cloud seeding is used minus the expected response under random control when all the other explanatory variables are the same.

\begin{figure}[t]
\begin{center}
\includegraphics[height=7.2 cm]{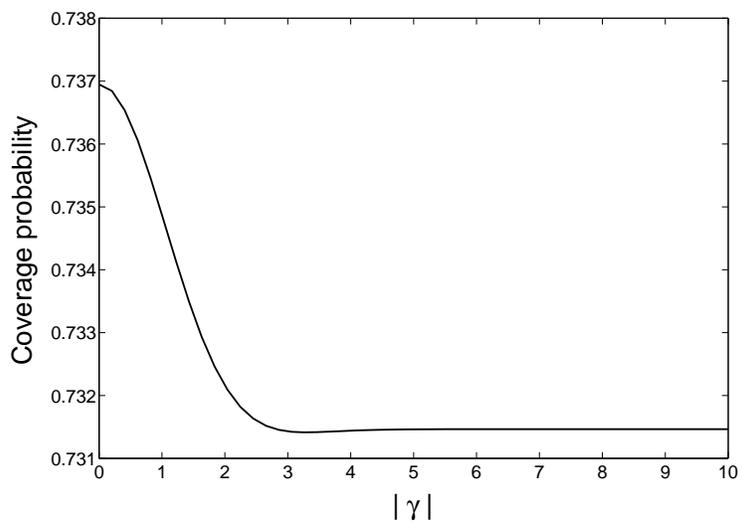}
\caption{\small{Plot of the coverage probability for the MPI, with nominal coverage $0.95$, for the seeding effect in the cloud seeding example when the submodel is defined by setting the coefficient of the squared seedability equal to zero.}}
\label{fig1}
\end{center}
\end{figure}

\begin{figure}[t]
\begin{center}
\includegraphics[height=7.2 cm]{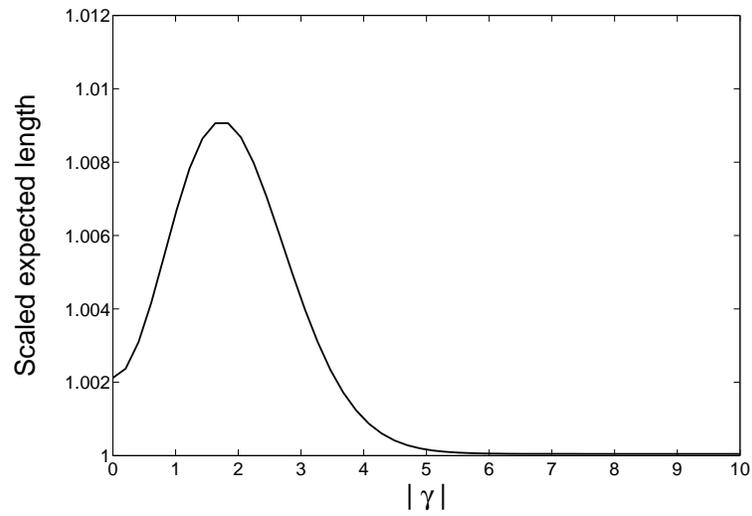}
\caption{\small{Plot of the scaled expected length for the MPI, with nominal coverage $0.95$, for the seeding effect in the cloud seeding example when the submodel is defined by setting the coefficient of the squared seedability equal to zero.}}
\label{fig2}
\end{center}
\end{figure}

We can construct profile-likelihood and MPI (over ${\cal M}_2$ and ${\cal M}_1$) confidence intervals for $\theta$.   The standard $0.95$ Student t confidence interval for $\theta$ under model ${\cal M}_2$ is $[-0.327, \, 3.421]$, the profile-likelihood confidence interval for $\theta$ under model ${\cal M}_2$, with nominal coverage $0.95$, is $[0.554,\, 2.539]$ and the MPI for $\theta$, with nominal coverage $0.95$, is $[0.618,\,2.572]$.  For comparison, the standard confidence interval for $\theta$, with nominal coverage $0.95$, after selection between models ${\cal M}_1$ and ${\cal M}_2$
using AIC but ignoring the model selection process is $[0.474,\, 2.650]$.  Model averaged profile likelihood confidence intervals for $\theta$ are held to be better than the confidence interval that ignores the model selection process, because they should better reflect the uncertainty in choosing between the two models.
For the MPI, we plot the exact coverage and the scaled expected length in Figures \ref{fig1} and \ref{fig2}, respectively.  We find that the coverage probability of the MPI is close to $0.7315$ for all $\gamma$ rather than the nominal $0.95$ and the scaled expected length is close to one for all $\gamma$.  Therefore, the MPI is actually similar to the standard $0.7315$ confidence interval for $\theta$.  This is not quite the good performance hoped for under model averaging.

\begin{center}
[Figures 1 and 2 near here]
\end{center}

\begin{figure}[t]
\begin{center}
\includegraphics[height=7.2 cm]{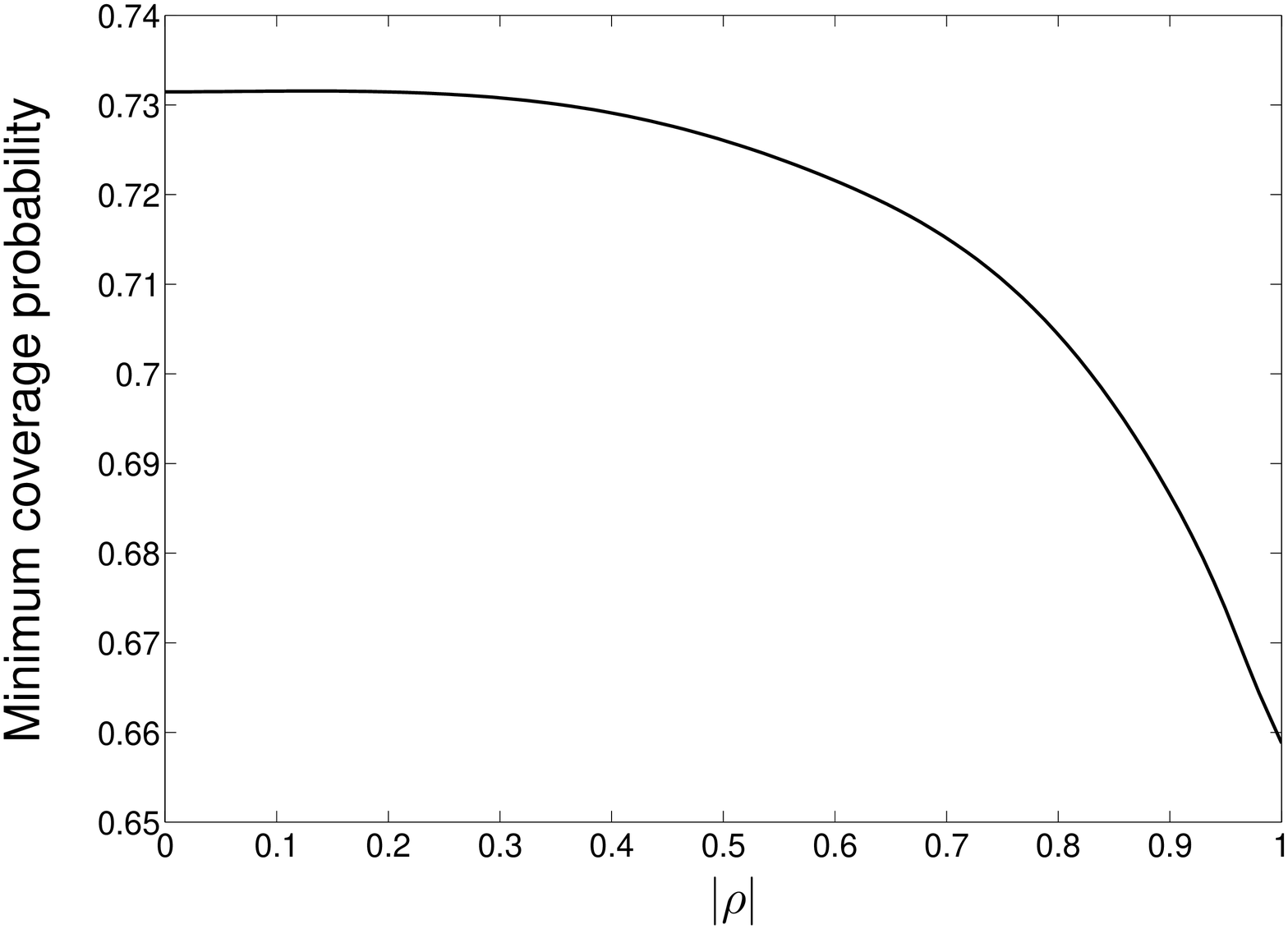}
\caption{\small{Plot of the minimum coverage against $|\rho|$ for the MPI, with nominal coverage $0.95$, for the seeding effect in the cloud seeding example when the submodel is defined by setting the coefficient of the squared seedability equal to zero.}}
\label{fig3}
\end{center}
\end{figure}

The important quantities for the MPI based on models ${\cal M}_2$ and ${\cal M}_1$ are $p/n$ and $\rho$, the correlation between the least squares estimators of $\theta$ and $\tau$.  For the cloud seeding example, $p/n=2/3$ which is not small and the correlation between $\hat{\theta}$ and $\hat{\tau}$ (which depends on $\XX$ and the choice of $\theta$ and $\tau$ so is known) is $\rho= 0.2472$ which is small and positive. The minimum coverage against $|\rho|$ for fixed $p/n = 2/3$ is plotted in Figure \ref{fig3}.

\begin{center}
[Figure 3 near here]
\end{center}

\begin{figure}[t]
\begin{center}
\includegraphics[height=7.2 cm]{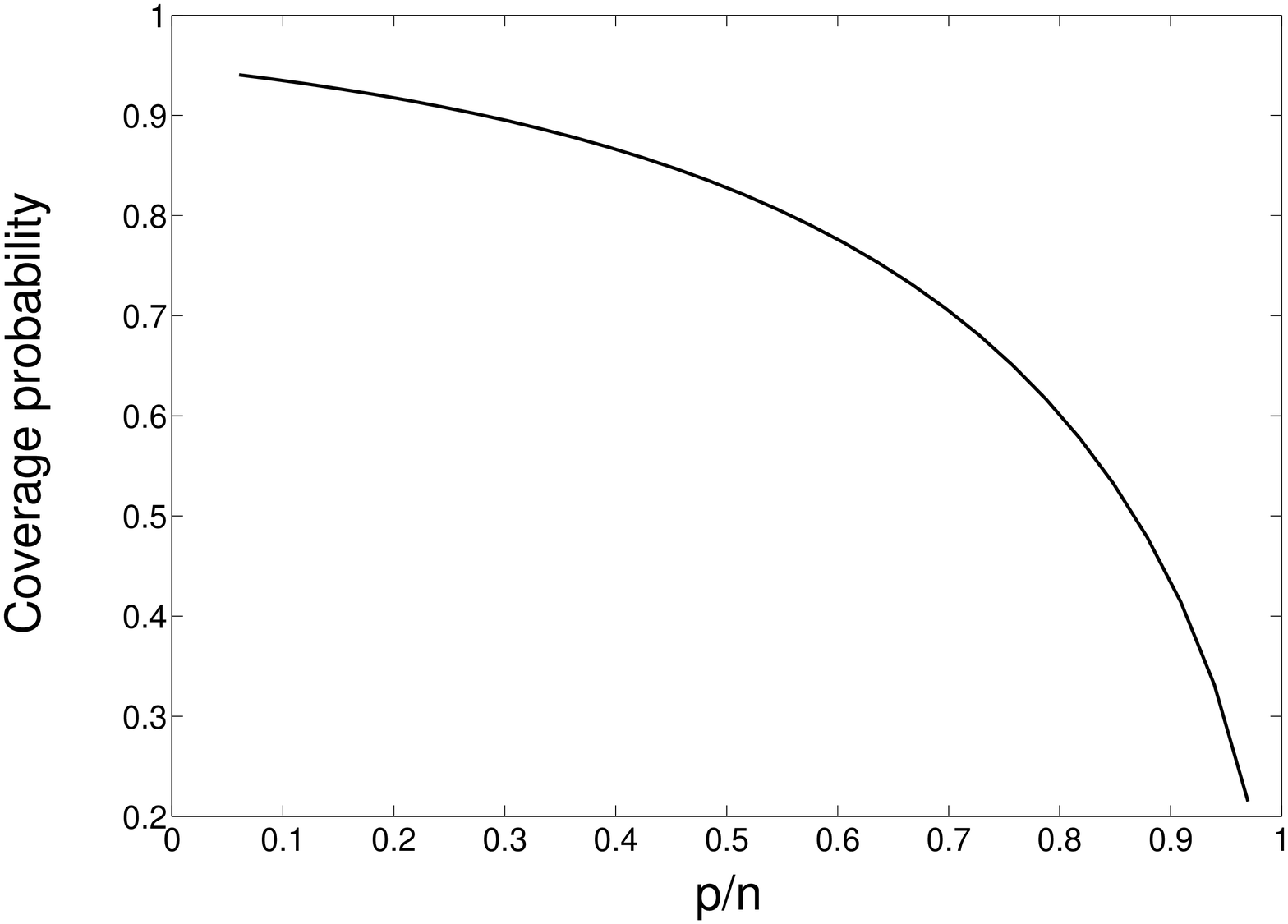}
\caption{\small{Plot of the coverage probability of the profile likelihood confidence interval under ${\cal M}_2$, with nominal coverage $0.95$, against $p/n$ when $n=33$.}}
\label{fig4}
\end{center}
\end{figure}

The coverage properties of the MPI as a function of $p/n$ are inherited from those of the profile likelihood confidence interval. We showed in Section \ref{sec:four} that the minimum coverage probability of the MPI, with nominal coverage $1-\alpha$, cannot be larger than the coverage probability of the profile likelihood confidence interval under ${\cal M}_2$, with the same nominal coverage. An asymptotic expansion of the latter coverage probability showed that it will be substantially below $1-\alpha$, unless $p/n$ is small (obviously, $0 < p/n < 1$).  This is confirmed by plotting the coverage of the profile likelihood confidence interval under ${\cal M}_2$ against $p/n$ in Figure \ref{fig4}.  The coverage decreases strongly as either $|\rho|$ or $p/n$ increase; in the cloud seeding example, the poor coverage is driven by $p/n$ not being small.

\begin{center}
[Figure 4 near here]
\end{center}

\begin{figure}[t]
\begin{center}
\includegraphics[height=7.2 cm]{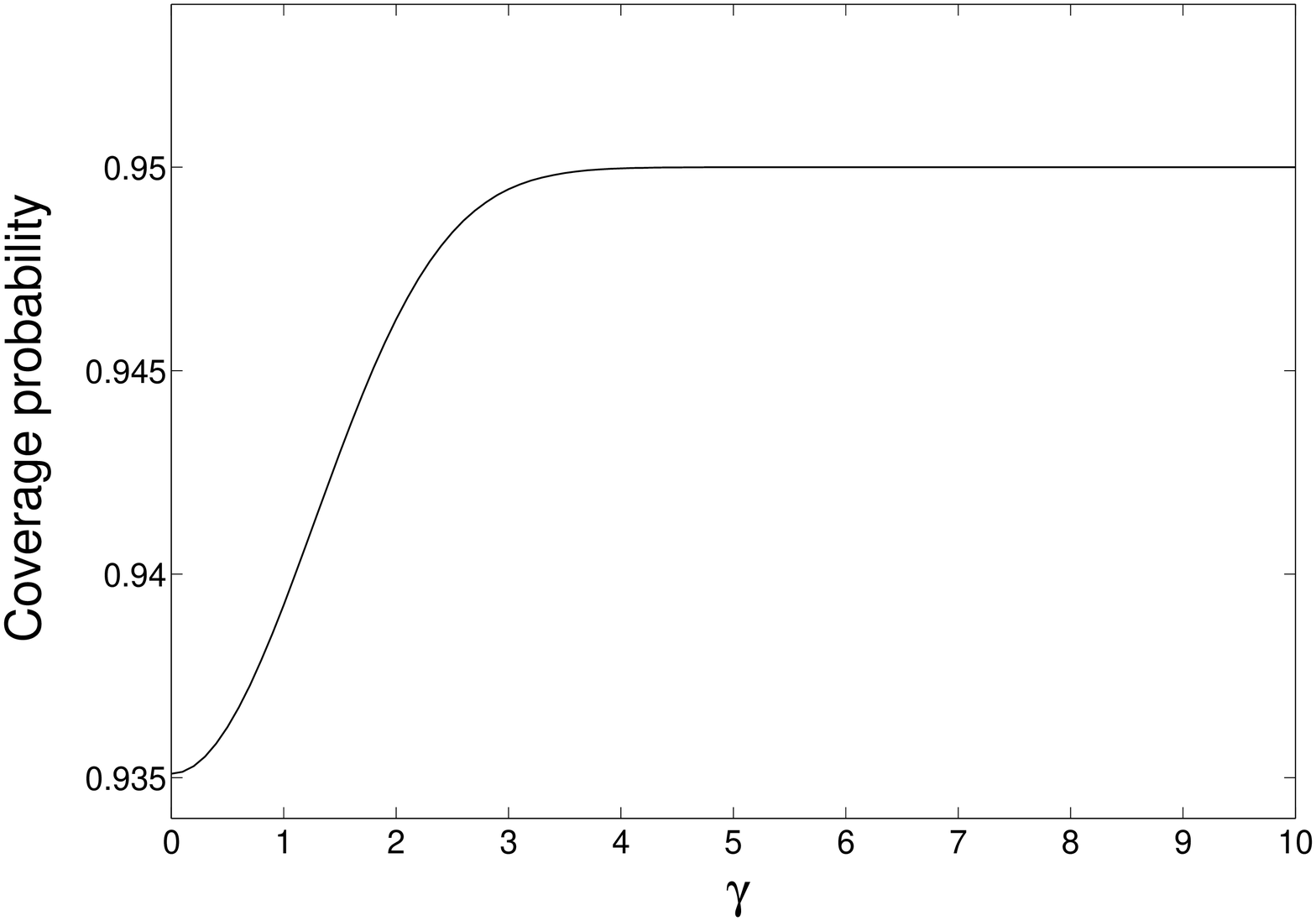}
\caption{\small{Plot of the coverage probability for the post-model-selection confidence interval, with nominal coverage $0.95$, for the seeding effect in the cloud seeding example when the possible models are the full model and the submodel defined by setting the coefficient of the squared seedability equal to zero. The model selected is the model with smaller AIC.}}
\label{fig5}
\end{center}
\end{figure}

It is interesting to compare the MPI interval with the naive confidence interval constructed after selecting between models ${\cal M}_1$ and ${\cal M}_2$ the model with smaller AIC and treating the selected model as if it had been given to us \textit{a priori} as the true model.  The coverage probability of this interval as a function of $\gamma$ is shown in Figure \ref{fig5} (Kabaila and Giri, 2009a, b).  Comparing this with Figure \ref{fig1}, we see that the coverage probability for this naive post-model-selection interval is uniformly far better than that of the MPI.

\begin{center}
[Figure 5 near here]
\end{center}

\begin{figure}[t]
\begin{center}
\includegraphics[height=7 cm]{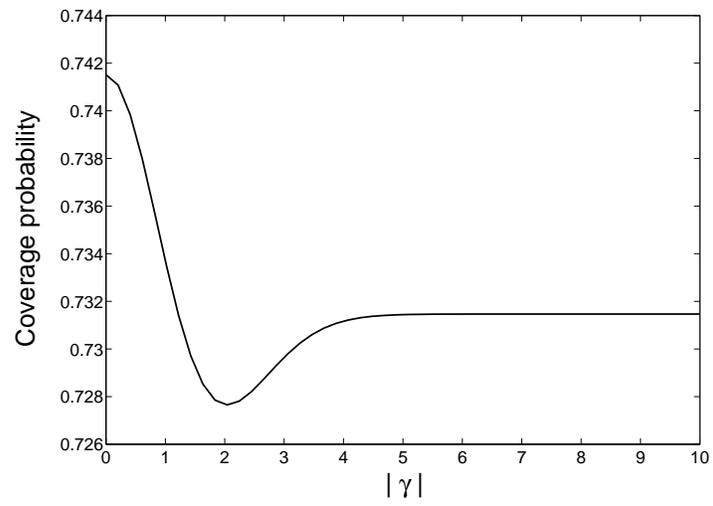}
\caption{\small{Plot of the coverage probability for the MPI, with nominal coverage $0.95$, for the seeding effect in the cloud seeding example when the submodel is defined by setting the coefficient of the seedability-earliness interaction equal to zero.}}
\label{fig6}
\end{center}
\end{figure}

\begin{figure}[t]
\begin{center}
\includegraphics[height=7 cm]{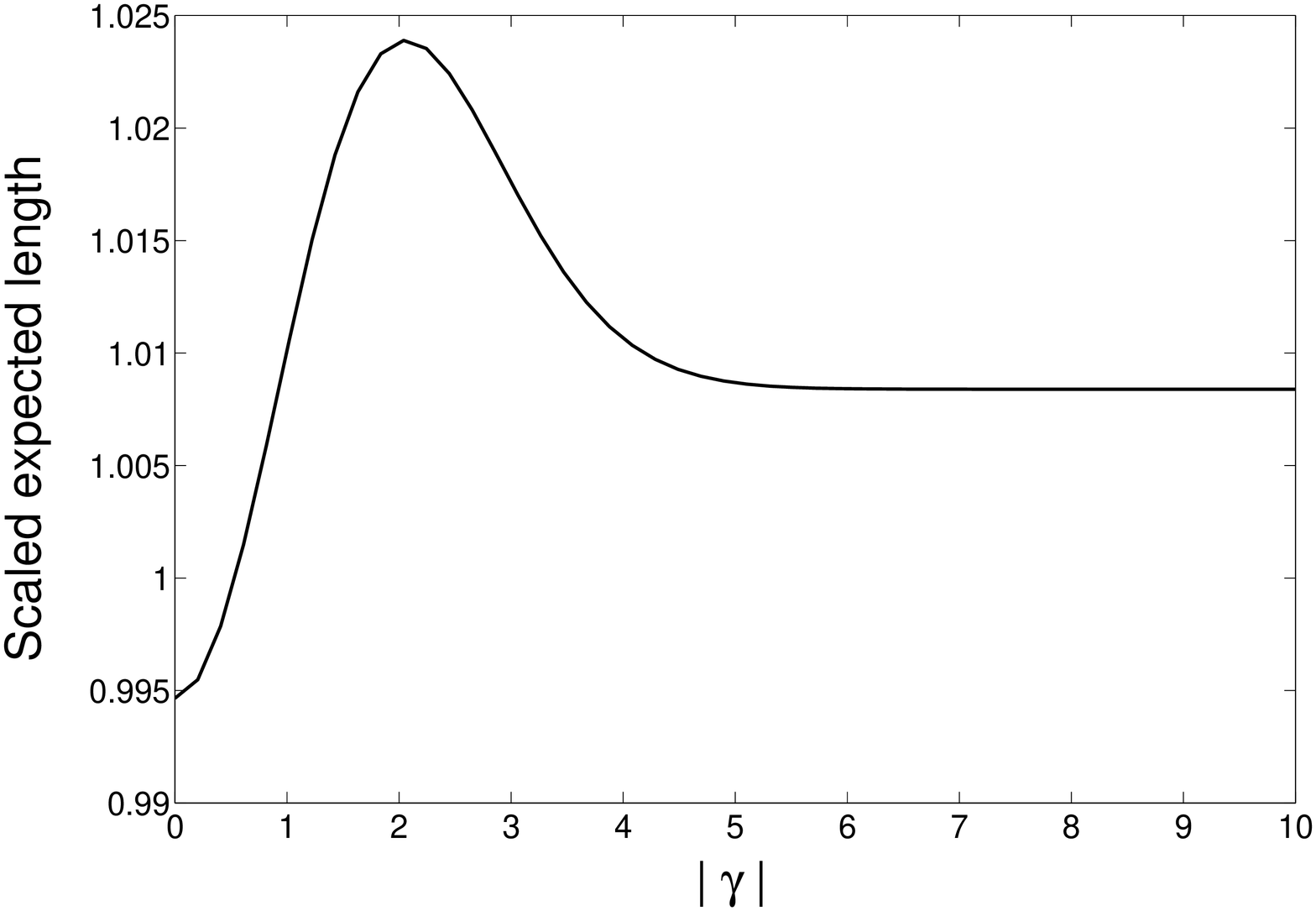}
\caption{\small{Plot of the scaled expected length for the MPI, with nominal coverage $0.95$, for the seeding effect in the cloud seeding example when the submodel is defined by setting the coefficient of the seedability-earliness interaction equal to zero.}}
\label{fig7}
\end{center}
\end{figure}

For a second example, suppose that we change $\tau$ from the coefficient of the squared seedability to the seedability-earliness interaction.  In this case, $n$ and $p$ are unchanged but now $\rho=-0.4530$.  The MPI for $\theta$, with nominal coverage $0.95$, is $[0.689,\,2.540]$, which is quite similar to the previous case.  We plot the exact coverage and the scaled expected length for the MPI in Figures \ref{fig6}
and \ref{fig7}, respectively.  The coverage probability of the MPI is close to $0.728$ for all $\gamma$ rather than the nominal $0.95$ and the scaled expected length is close to one for all $\gamma$, although the curves are different from those obtained in Figures \ref{fig1} and
\ref{fig2}.  We conclude that the MPI has similar coverage and expected length properties to the standard $0.728$ confidence interval for $\theta$.  The naive confidence interval constructed after selecting between models ${\cal M}_1$ and ${\cal M}_2$ the model with smaller AIC and treating the selected model as if it had been given to us \textit{a priori} as the true model has similar coverage to that shown in Figure \ref{fig5}.  Once again, we see that the coverage probability for the naive post-model-selection interval is uniformly far better than that of the MPI.

\begin{center}
[Figures 6 and 7 near here]
\end{center}

\section{Conclusion}

We have examined the exact coverage and scaled expected length of the MPI for a parameter $\theta$, with nominal coverage $1-\alpha$, in a particular simple situation in which there are two linear regression models (differing in only a single parameter $\tau$) to average over.  We showed that both the coverage and the scaled expected length depend on $n$, $n-p$, the correlation $\rho$ between the least squares estimators $\hat{\theta}$ and $\hat{\tau}$, and the unknown true value $\gamma=\tau \big/ \big(\sigma v_{\tau}^{1/2} \big)$. As $\gamma$ is unknown, it is useful to consider the minimum coverage and the maximum scaled expected length over $\gamma$.  The results show that the MPI can perform poorly when $p/n$ is not small or when $|\rho|$ is large, and should not be used in these situations.  In fact, in these situations, the MPI performs no better than than post model selection confidence intervals which ignore the selection process.

The MPI is obtained by trying to average profile likelihood confidence intervals and we have shown that the performance of the MPI is limited by the performance of the underlying profile likelihood confidence intervals.  In particular, the MPI inherits poor performance when $p/n$ is not small from the fact that profile likelihood confidence intervals perform poorly when $p/n$ is not small.  Averaging other types of confidence intervals which do not have this problem may lead to better results, at least when $p/n$ is not small.

\section*{References}

\smallskip

\rf Biondini, R., Simpson, J. and Woodley, W. (1977). Empirical predictors for natural and seeded rainfall in the Florida area cumulus experiment (FACE), 1970-1975. \textit{Journal of Applied Meteorology} \textbf{16}, 585--594.

\smallskip

\rf Buckland, S.T., Burnham, K.P. and Augustin, N.H. (1997). Model selection: an integral
part of inference. \textit{Biometrics} \textbf{53}, 603--618.

\smallskip

\rf Burnham, K.P. and Anderson, D.R. (2002). {\sl Model selection and multimodel
inference, a practical information-theoretic approach}, 2nd edition. Springer: New York.

\smallskip

\rf Claeskens, G. and Hjort, N.L. (2008). {\sl Model selection and model averaging}.
Cambridge University Press.

%
%

\smallskip

\rf Fletcher, D. and Dillingham, P.W. (2011). Model-averaged confidence intervals for factorial experiments.
\textit{Comput. Statist. Data Anal.} \textbf{55}, 3041--3048.

\smallskip

\rf Fletcher, D. and Turek, D. (2011). Model-averaged profile likelihood intervals. \textit{J. Agric. Biol. Environ. Stat.} \textbf{17}, 38--51.

\smallskip

\rf Hoeting, J., Madigan, D., Raftery, A.E. and Volinsky, C.T. (1999). Bayesian model-averaging: A tutorial (with discussion). \textit{Statist. Sci.} \textbf{14},
382--417.

\smallskip

\rf Hjort, N.L. and Claeskens, G. (2003). Frequentist model average estimators. \textit{J. Amer. Statist. Assoc.}
\textbf{98}, 879--899.

\smallskip

\rf Kabaila, P. (2005). On the coverage probability of confidence intervals in regression after variable selection. \textit{Aust. N. Z. J. Stat.} \textbf{47}, 549--562.

\smallskip

\rf Kabaila, P. (2009). The coverage properties of confidence regions after model selection.
\textit{International Statistical Review} \textbf{77}, 405--414.

\smallskip

\rf Kabaila, P. and Leeb, H. (2006). On the large-sample minimal coverage probability of confidence intervals
after model selection. \textit{J. Amer. Statist. Assoc.} \textbf{101}, 619--629.

\smallskip

\rf Kabaila, P. and Giri, K. (2009a). Upper bounds on the minimum coverage probability
of confidence intervals in regression after model selection.
\textit{Aust. N. Z. J. Stat.} \textbf{51}, 271--287.

\smallskip

\rf Kabaila, P. and Giri, K. (2009b). Confidence intervals in regression utilizing uncertain prior information. \textit{J. Statist. Plann. Inference} \textbf{139}, 3419--3429.

%
%

\smallskip

\rf Lukacs, P.M., Burnham, K.P. and Anderson, D.R. (2010).  Model selection bias and Freedman's paradox.  \textit{Ann. Inst. Statist. Math.} \textbf{62}, 117--125.

\smallskip

\rf Miller, A. (2002). \textit{Subset selection in regression}, 2nd edition. Chapman \& Hall/CRC.

%
%

\smallskip

\rf Raftery, A.E., Madigan, D. and Hoeting, J.A. (1997). Bayesian model-averaging for linear regression models. \textit{J. Amer. Statist. Assoc.} \textbf{92},
179--191.

\smallskip

\rf Turek, D. and Fletcher, D. (2012). Model-averaged Wald confidence intervals. \textit{Comput. Statist. Data Anal.} \textbf{56},
2809--2815.

\smallskip

\rf Volinsky, C.T., Madigan, D., Raftery, A.E. and Kronmal, R.A. (1997). Bayesian model-averaging in proportional hazard models: Assessing the risk of a stroke. \textit{J. R. Statist. Soc. Ser. C Appl. Stat.} \textbf{46},
433--448.

\smallskip

\rf Wang, H. and Zou, S.Z.F. (2013). Interval estimation by frequentist model averaging.
\textit{Commun. Statist. Theory Methods} to appear.

\bigskip \noindent
Paul Kabaila,
Department of Mathematics and Statistics,
La Trobe University, Victoria 3086, Australia.\\
E-mail: P.Kabaila@latrobe.edu.au

\section*{Appendix}

\subsection*{The models}

It simplifies the presentation if we reparametrise the models ${\cal M}_1$ and ${\cal M}_2$ to be explicit functions of the parameters $\theta$ and $\tau$.  Let $\MM$ be the $p\times p$ matrix with first two rows given by $\ba^{\top}(\XX^{\top}\XX)^{-1/2}$ and $\cc^{\top}(\XX^{\top}\XX)^{-1/2}$, respectively, and the remaining $p-2$ rows given by orthonormal $p$-vectors that are orthogonal to both $\ba$ and $\cc$.  The incorporation of $(\XX^{\top}\XX)^{-1/2}$ into the first two rows of $\MM$ may seem unnecessary but in fact, as we will see below, it produces a useful standardisation.  The model ${\cal M}_2$ can be written as
\[
\tilde{\YY} = \tilde{\XX} \et + \vare,
\]
where $\tilde{\YY} = \YY - t\XX(\XX^{\top}\XX)^{-1/2}\MM^{-1}\ee_2$, $\tilde{\XX}=\XX(\XX^{\top}\XX)^{-1/2}\MM^{-1}$ and $\et = \MM(\XX^{\top}\XX)^{1/2}\be - t\ee_2$, with $\ee_2$ a $p$-vector with the second component equal to one and all other components equal to zero.  Write $\et = (\theta, \tau, \ps^{\top})^{\top}$, where $\ps$ is the $(p-2)$-vector of the remaining regression parameters. The model ${\cal M}_1$ is ${\cal M}_2$ with $\tau=0$.

\subsection*{The likelihood for the models}

We can write down the log-likelihood for the reparametrised model directly and then re-express it in terms of the maximum likelihood estimators of the parameters, which are a minimal sufficient statistic for ${\cal M}_2$ and ${\cal M}_1$. It is simpler to first reduce the data and work from the sampling distribution of this minimal sufficient statistic. The maximum likelihood estimator of $\et$ is given by
\[
\hat{\et} = \big(\tilde{\XX}^{\top}\tilde{\XX} \big)^{-1} \tilde{\XX}^{\top} \tilde{\YY}
= \MM(\XX^{\top}\XX)^{-1/2}\XX^{\top}\YY  - t\ee_2
\]
and the maximum likelihood estimator of $\sigma^2$ is $(n-p)\hat{\sigma}^2/n$, where
\[
\hat{\sigma}^2 = (\tilde{\YY} - \tilde{\XX} \hat{\et})^{\top}(\tilde{\YY} - \tilde{\XX} \hat{\et})/(n-p) = (\YY - \XX\hat{\be})^{\top}(\YY - \XX\hat{\be})/(n-p).
\]
We have
\[
\hat{\et} \sim N \left(\left[\begin{array}{c} \theta\\ \tau\\ \ps\end{array}\right], \, \sigma^2 \left[\begin{array}{ccc} v_{\theta} & \rho (v_{\theta}v_{\tau})^{1/2} & \zer^{\top} \\ \rho(v_{\theta}v_{\tau})^{1/2} & v_{\tau} & \zer^{\top}\\ \zer & \zer & \II_{p-2}\end{array},\right]\right),
\]
where $v_{\theta} = \ba^{\top}(\XX^{\top}\XX)^{-1}\ba$, $v_{\tau} = \cc^{\top}(\XX^{\top}\XX)^{-1}\cc$ and $\rho=\ba^{\top}(\XX^{\top}\XX)^{-1}\cc/(v_{\theta}v_{\tau})^{1/2}$ are known quantities, and, independently,
\[
(n-p)\hat{\sigma}^2/\sigma^2 \sim \chi^2_{n-p}.
\]
The advantage of incorporating $(\XX^{\top}\XX)^{-1/2}$ into the first two rows of $\MM$ is that the sampling distribution of $\hat{\ps}$ has a very simple covariance structure with unknown parameter $\sigma^2$.
We can write down the log-likelihood for ${\cal M}_2$ which (discarding terms which do not depend on the unknown parameters) is
\begin{eqnarray*}
\ell_2(\theta, \tau, \ps, \sigma^2)
&=& -\frac{n}{2}\log(\sigma^2) - \frac{1}{2\sigma^2}\Bigg[\frac{1}{(1-\rho^2)} \Bigg\{\frac{(\hat{\theta} - \theta)^2}{v_{\theta}} + \frac{(\hat{\tau} - \tau)^2}{v_{\tau}} - 2\rho\frac{(\hat{\theta} - \theta)(\hat{\tau} - \tau)}{v_{\theta}^{1/2}v_{\tau}^{1/2}} \Bigg\} \\
&& + (\hat{\ps}-\ps)^T(\hat{\ps}-\ps) + (n-p) \hat{\sigma}^2 \Bigg],
\end{eqnarray*}
and hence the log-likelihood for ${\cal M}_1$ is $\ell_1(\theta, \ps, \sigma^2) = \ell_2(\theta, 0, \ps, \sigma^2)$.

We do not have to specify the particular underlying linear regression model or the specific parameters $\theta$ and $\tau$.  The results below hold for any full-rank linear regression model and for any linear combinations $\theta$ and $\tau$ of the regression parameter $\be$.

\subsubsection*{The signed-root log-likelihood statistic for ${\cal M}_2$}

Setting the derivatives of the log-likelihood $\ell_2(\theta, \tau, \ps, \sigma^2)$ with respect to the unknown parameters to zero and solving the resulting estimating equations shows that the maximum likelihood estimators are $\hat{\theta}$, $\hat{\tau}$, $\hat{\ps}$ and $(n-p) \hat{\sigma}^2/n$, respectively, so the maximum value of the log-likelihood is
\begin{eqnarray*}
\ell_2 \left(\hat{\theta}, \hat{\tau}, \hat{\ps}, (n-p) \hat{\sigma}^2/n \right)
&=& -\frac{n}{2}\log \big\{(n-p) \hat{\sigma}^2/n \big\} - \frac{n}{2}.
\end{eqnarray*}
Next, holding $\theta$ fixed and setting the derivatives of the log-likelihood $\ell_2(\theta, \tau, \ps, \sigma^2)$ with respect to the remaining unknown parameters to zero, we obtain the 
maximum profile likelihood estimators
$\hat{\tau}(\theta) = \hat{\tau} - \rho(v_{\tau}/v_{\theta})^{1/2}(\hat{\theta} - \theta)$, $\hat{\ps}(\theta)=\hat{\ps}$ and $\hat{\sigma}^2(\theta) = \{(\hat{\theta} - \theta)^2/v_{\theta} + (n-p) \hat{\sigma}^2 \}/n$, so the maximum value of the profile log-likelihood is
\begin{eqnarray*}
\ell_2 \left\{\theta, \hat{\tau}(\theta), \hat{\ps}(\theta), \hat{\sigma}^2(\theta) \right\}
&=& -\frac{n}{2}\log \left[\left\{\frac{(\hat{\theta} - \theta)^2}{v_{\theta}}+ (n-p) \hat{\sigma}^2 \right\} \Big/n \right] - \frac{n}{2}
\end{eqnarray*}
It follows that the signed root log-likelihood ratio statistic for ${\cal M}_2$ is $r_2 \big\{(\hat{\theta}-\theta)/v_{\theta}^{1/2}, \, \hat{\sigma} \big\} = r_2 \big\{(\hat{\theta}-\theta)/(\sigma v_{\theta}^{1/2}), \, \hat{\sigma}/\sigma \big\}$, where $r_2$ is given by (\ref{eq:r2}).

\subsubsection*{The signed-root log-likelihood statistic for ${\cal M}_1$}

The log-likelihood for model ${\cal M}_1$ is
\begin{eqnarray*}
\ell_1(\theta, \ps, \sigma^2)
&=& -\frac{n}{2}\log(\sigma^2) - \frac{1}{2\sigma^2}\Bigg[\frac{1}{(1-\rho^2)} \Bigg\{\frac{(\hat{\theta} - \theta)^2}{v_{\theta}} + \frac{\hat{\tau}^2}{v_{\tau}} - 2\rho\frac{(\hat{\theta} - \theta)\hat{\tau}}{v_{\theta}^{1/2}v_{\tau}^{1/2}} \Bigg\} \\
&& + (\hat{\ps}-\ps)^T(\hat{\ps}-\ps) + (n-p) \hat{\sigma}^2 \Bigg]
\end{eqnarray*}
Setting the derivatives of the log-likelihood with respect to the unknown parameters to zero and solving the resulting estimating equations shows that the maximum likelihood estimators are $\hat{\theta}-\rho(v_{\theta}/v_{\tau})^{1/2}\hat{\tau}$, $\hat{\ps}$ and $\{\hat{\tau}^2/v_{\tau} + (n-p) \hat{\sigma}^2 \}/n$, respectively, so the maximum value of the log-likelihood is
\begin{eqnarray*}
\ell_1 \left(\hat{\theta}-\rho(v_{\theta}/v_{\tau})^{1/2}\hat{\tau}, \hat{\ps}, \{\hat{\tau}^2/v_{\tau} + (n-p) \hat{\sigma}^2 \}/n \right)
&=& -\frac{n}{2}\log \big[\{\hat{\tau}^2/v_{\tau} + (n-p) \hat{\sigma}^2 \}/n \big] - \frac{n}{2}.
\end{eqnarray*}
Next, holding $\theta$ fixed and setting the derivatives of the log-likelihood $\ell_1(\theta, \ps, \sigma^2)$ with respect to the remaining unknown parameters to zero, we obtain the 
maximum profile likelihood estimators
$\hat{\ps}_1(\theta)
=\hat{\ps}$ and
$\hat{\sigma}^2(\theta) = \Big[\frac{1}{(1-\rho^2)}\{(\hat{\theta} - \theta)^2/v_{\theta} + \hat{\tau}^2/v_{\tau} - 2\rho(\hat{\theta} - \theta)\hat{\tau}/(v_{\theta}v_{\tau})^{1/2}\}+ (n-p) \hat{\sigma}^2 \Big] \big/n$, so the maximum value of the profile log-likelihood is
\begin{eqnarray*}
\lefteqn{\ell_1 \big\{\theta, \hat{\ps}_1, \hat{\sigma}_1^2(\theta) \big\}}\\
 &=& -\frac{n}{2}\log\Bigg(\Big[\frac{1}{(1-\rho^2)}\Big\{\frac{(\hat{\theta} - \theta)^2}{v_{\theta}} + \frac{\hat{\tau}^2}{v_{\tau}} - 2\rho\frac{(\hat{\theta} - \theta)\hat{\tau}}{v_{\theta}^{1/2}v_{\tau}^{1/2}} \Big\}+ (n-p) \hat{\sigma}^2 \Big]/n\Bigg) - \frac{n}{2}\\
&=& -\frac{n}{2}\log\Big(\Big[\frac{1}{(1-\rho^2)}\Big\{\frac{(\hat{\theta} - \theta)^2}{v_{\theta}} + \rho^2\frac{\hat{\tau}^2}{v_{\tau}} - 2\rho\frac{(\hat{\theta} - \theta)\hat{\tau}}{v_{\theta}^{1/2}v_{\tau}^{1/2}}\Big\}  + \frac{\hat{\tau}^2}{v_{\tau}}+ (n-p) \hat{\sigma}^2 \Big]/n\Big) - \frac{n}{2}.
\end{eqnarray*}
It follows that the signed root log-likelihood ratio statistic for ${\cal M}_1$ is
$r_1 \big\{(\hat{\theta}-\theta)\big/\big(\sigma v_{\theta}^{1/2}\big), \hat{\tau}/(\sigma v_{\tau}^{1/2}), \hat{\sigma}/\sigma \big\}$,
where $r_1$ is given by (\ref{eq:r1}).

\subsection*{Akaike weights}

For $d=2$, the Akaike Information Criteria (AIC) for the two models are
\[
\mbox{AIC}_2 = n\log\{(n-p) \hat{\sigma}^2/n\} + dp
\]
and
\[
\mbox{AIC}_1 = n\log[\{(\hat{\tau}^2/v_{\tau}) + (n-p) \hat{\sigma}^2 \}/n] + d(p-1),
\]
respectively, so the weight is
\begin{eqnarray*}
w_1 &=& \frac{\exp\big\{-\frac{1}{2}(\mbox{AIC}_1 - \mbox{AIC}_{\min})\big\}}{\exp\big\{-\frac{1}{2}(\mbox{AIC}_1 - \mbox{AIC}_{\min})\big\} + \exp\big\{-\frac{1}{2}(\mbox{AIC}_2 - \mbox{AIC}_{\min})\big\}} \\
&=& \frac{1}{1 + \exp\big\{\frac{1}{2}(\mbox{AIC}_1 - \mbox{AIC}_2)\big\}}\\
&=& \frac{1}{1 + \Big\{1+\frac{\hat{\tau}^2}{(n-p) \hat{\sigma}^2 v_{\tau}}\Big\}^{n/2} \exp(-d/2)}.
\end{eqnarray*}
This corresponds to the expression (\ref{eq:weight}).

We can calibrate the choice of $d$ by considering the hypothesis test in which we reject model ${\cal M}_1$ in favour of ${\cal M}_2$ when $\mbox{AIC}_2 < \mbox{AIC}_1$.  When model ${\cal M}_1$ is true, the probability of rejecting ${\cal M}_1$ (i.e. the level of the test) is
\[
2\left(1-G_{n-p}\left[(n-p)^{1/2}\left\{\exp\left(\frac{d}{n}\right) - 1\right\}^{1/2}\right]\right),
\]
where $G_{n-p}$ is the cumulative distribution function of the Student t distribution with $n-p$ degrees of freedom.  If we set the level of the test equal to $u$, we find
\begin{eqnarray*}
d &=& n \log\left[1 + \frac{\{G_{n-p}^{-1}(1-u/2)\}^2}{n-p}\right].
\end{eqnarray*}
Expanding the $\log$ function and then letting $n\rightarrow \infty$, we find that
\begin{eqnarray*}
d &=& \big\{G_{n-p}^{-1}(1-u/2) \big\}^2 + O(n^{-1}) \rightarrow z_{1-u/2}^2, 
\end{eqnarray*}
which can also be expressed in terms of the chi-squared distribution with one degree of freedom.

\subsection*{Proof of Theorem 1}

The coverage probability of the MPI confidence interval, with nominal coverage $1-\alpha$,  is
\[
P \left(\hat{\theta}_l \le \theta \le \hat{\theta}_u \right)
= 1- P \left(\theta < \hat{\theta}_l \right)  - P \left(\hat{\theta}_u > \theta \right).
\]
Now $h(\delta,x,y)$ is an increasing function of $\delta$ for fixed $x$ and $y$ so
\begin{eqnarray*}
\lefteqn{P \left(\theta < \hat{\theta}_l \right)}\\
&=& P\left\{(\hat{\theta} - \theta)\big/\big(\sigma v_{\theta}^{1/2}\big) > (\hat{\theta}-\hat{\theta}_l)\big/\big(\sigma v_{\theta}^{1/2}\big) \right\} \\
&=& P\left[h\left\{(\hat{\theta}-\theta)\big/\big(\sigma v_{\theta}^{1/2}\big), \hat{\tau}\big/\big(\sigma v_{\tau}^{1/2}\big), \hat{\sigma}/\sigma \right\} > 1-\alpha/2 \right]\\
 &=& P \left[(\hat{\theta}-\theta)/(\sigma v_{\theta}^{1/2}) > \delta_{1-\alpha/2}\big\{\hat{\tau}/(\sigma v_{\tau}^{1/2}), \hat{\sigma}/\sigma \big\}\right]\\
  &=& \int_{0}^{\infty}\int_{-\infty}^{\infty}P[(\hat{\theta}-\theta)/(\sigma v_{\theta}^{1/2}) > \delta_{1-\alpha/2}\{\hat{\tau}/(\sigma v_{\tau}^{1/2}), \hat{\sigma}/\sigma\}\big|\hat{\tau}/(\sigma v_{\tau}^{1/2}) = x, \hat{\sigma}/\sigma=y]\\
  && \times \phi(x-\gamma) \, f_{n-p}(y) \, dx \, dy,
\end{eqnarray*}
where $\gamma=\tau/(\sigma v_{\tau}^{1/2})$.  Now the distribution of
$(\hat{\theta}-\theta)/(\sigma v_{\theta}^{1/2})$
conditional on
$\hat{\tau}/(\sigma v_{\tau}^{1/2}) = x$ is $N\big(\rho(x - \gamma), 1-\rho^2 \big)$, $\hat{\tau}/(\sigma v_{\tau}^{1/2}) \sim N(\gamma, 1)$ and $\hat{\theta}$ and $\hat{\tau}$ are  independent of $\hat{\sigma}$, so
\begin{eqnarray*}
\lefteqn{P \big[(\hat{\theta}-\theta)/(\sigma v_{\theta}^{1/2}) > \delta_{1-\alpha/2}\{\hat{\tau}/(\sigma v_{\tau}^{1/2}), \hat{\sigma}/\sigma\}\, \big| \, \hat{\tau}/(\sigma v_{\tau}^{1/2}) = x, \, \hat{\sigma}/\sigma=y \big]}\\
 &=& P \big\{(\hat{\theta}-\theta)/(\sigma v_{\theta}^{1/2}) > \delta_{1-\alpha/2}(x, y)\, \big|\,
 \hat{\tau}/(\sigma v_{\tau}^{1/2}) = x \big\} \\
 &=& 1- P \big\{(\hat{\theta}-\theta)/\sigma v_{\theta}^{1/2} \le \delta_{1-\alpha/2}(x, y)\, \big| \,
 \hat{\tau}/(\sigma v_{\tau}^{1/2})= x \big\}\\
&=& 1- \Phi \left\{\frac{\delta_{1-\alpha/2}(x, y) - \rho(x-\gamma)}{(1-\rho^2)^{1/2}} \right\}\\
\end{eqnarray*}
and hence
\begin{eqnarray*}
P \left(\theta < \hat{\theta}_l \right)
&=& \int_{0}^{\infty}\int_{-\infty}^{\infty}\left[1- \Phi\left\{\frac{\delta_{1-\alpha/2}(x, y) - \rho(x-\gamma)}{(1-\rho^2)^{1/2}}\right\} \right]\phi(x-\gamma) f_{n-p}(y) \, dx \, dy.
\end{eqnarray*}
Similarly,
\begin{eqnarray*}
1-P \left(\hat{\theta}_u > \theta \right)
&=& P \left(\theta < \hat{\theta}_u \right)\\
 &=& P \left[h\{(\hat{\theta}-\theta)/(\sigma v_{\theta}^{1/2}), \hat{\tau}/(\sigma v_{\tau}^{1/2}), \hat{\sigma}/\sigma\} > \alpha/2 \right]\\
 &=& P \left[(\hat{\theta}-\theta)/(\sigma v_{\theta}^{1/2}) > \delta_{\alpha/2}\{\hat{\tau}/(\sigma v_{\tau}^{1/2}), \hat{\sigma}/\sigma\} \right]\\
&=& \int_{0}^{\infty}\int_{-\infty}^{\infty}\left[1- \Phi \left\{\frac{\delta_{\alpha/2}(x, y) - \rho(x-\gamma)}{(1-\rho^2)^{1/2}} \right\}\right] \phi(x-\gamma)\, f_{n-p}(y) \, dx \, dy.
\end{eqnarray*}

\subsection*{Proof of Corollary 1}

From the proof of Theorem 1, we can write
\begin{equation*}
P\big(\theta < \hat{\theta}_l \big)
= 1 - P \big\{h(G,H,W) \le 1 - \alpha/2 \big\},
\end{equation*}
where $h$ is defined in (\ref{eq:h}), $G = (\hat{\theta}  - \theta)/\sigma v_{\theta}^{1/2}\sim N(0,1)$, $H =\hat{\tau}/\sigma v_{\tau}^{1/2} \sim N(\gamma, 1)$,
$(n-p) W^2 = (n-p)\hat{\sigma}/\sigma\sim \chi^2_{n-p}$ and $(G,H)$ and $W$ are independent.
From (\ref{eq:weight}), $w_1(H,W)$ converges in probability to 0, as $\gamma \rightarrow \infty$.
Since $0 < \Phi(x) < 1$ for all $x \in \mathbb{R}$, this implies by (\ref{eq:h}) that $h(G,H,W)$ converges in
probability to $\Phi\{r_2(G,W)\}$, as $\gamma \rightarrow \infty$. Thus, $h(G,H,W)$ converges in
distribution to $\Phi\{r_2(G,W)\}$, as $\gamma \rightarrow \infty$. The cumulative distribution
function of $\Phi\{r_2(G,W)\}$, evaluated at $u$, is a continuous function of $u \in \mathbb{R}$. Therefore
\begin{equation*}
P\big(\theta < \hat{\theta}_l \big) \rightarrow 1 - P \big[\Phi\{r_2(G,W)\} \le 1 - \alpha/2 \big], \qquad \mbox { as } \gamma \rightarrow \infty.
\end{equation*}

Now consider the profile likelihood interval under ${\cal M}_2$, with nominal coverage $1-\alpha$.
The lower endpoint of this confidence interval, denoted by $\hat{\theta}_{2 l}$, is obtained by
solving for $\theta < \hat{\theta}$ in
\begin{equation*}
\Phi \left\{ r_2 \left(\frac{\hat{\theta}  - \theta}{\sigma v_{\theta}^{1/2}},  \frac{\hat{\sigma}}{\sigma} \right)\right\}
= 1 - \alpha/2.
\end{equation*}
Note that
\begin{align*}
P\big(\theta < \hat{\theta}_{2l} \big)
&= P \left(\frac{\hat{\theta}  - \theta}{\sigma v_{\theta}^{1/2}}
> \frac{\hat{\theta}  - \hat{\theta}_{2 l}}{\sigma v_{\theta}^{1/2}} \right) \\
&= P \left[\Phi \left\{r_2 \left(\frac{\hat{\theta}  - \theta}{\sigma v_{\theta}^{1/2}},  \frac{\hat{\sigma}}{\sigma} \right) \right\}
> \Phi \left\{r_2 \left(\frac{\hat{\theta}  - \hat{\theta}_{2 l}}{\sigma v_{\theta}^{1/2}},  \frac{\hat{\sigma}}{\sigma} \right) \right\} \right] \\
&= P \left[\Phi \big\{r_2(G,W) \big\} > 1 - \alpha/2 \right] \\
&= 1 - P \left[\Phi \big\{r_2(G,W) \big\} \le 1 - \alpha/2 \right]
\end{align*}
so $P\big(\theta < \hat{\theta}_l \big) \rightarrow P\big(\theta < \hat{\theta}_{2l} \big)$, as $\gamma \rightarrow \infty$.
Similarly, $P\big(\theta < \hat{\theta}_u \big) \rightarrow P\big(\theta < \hat{\theta}_{2u} \big)$, as $\gamma \rightarrow \infty$ and the first part of Corollary 1 holds.  The second part follows from showing that
\[
\hat{\theta}_{2l}
= \hat{\theta} - (n-p)^{1/2} \, \hat{\sigma} \, v_{\theta}^{1/2} \left\{\exp \left(z_{1-\alpha/2}^2 \big/n \right) -1 \right\}^{1/2}
\]
and
\[
\hat{\theta}_{2u}
= \hat{\theta} + (n-p)^{1/2} \, \hat{\sigma} \, v_{\theta}^{1/2} \left\{\exp \left(z_{1-\alpha/2}^2 \big/n \right) -1 \right\}^{1/2},
\]
where
$z_{1-\alpha/2} = \Phi^{-1}(1-\alpha/2)$, and calculating the coverage probability of the profile likelihood confidence interval with nominal coverage $1-\alpha$.

\subsection*{Proof of Theorem 2}

The expected length of the MPI confidence interval, with nominal coverage $1-\alpha$, is
\begin{eqnarray*}
\E \left(\hat{\theta}_u - \hat{\theta}_l \right)
&=& \sigma \, v_{\theta}^{1/2} \E \left\{(\hat{\theta} - \hat{\theta}_l)/(\sigma v_{\theta}^{1/2}) - (\hat{\theta} - \hat{\theta}_u)/(\sigma v_{\theta}^{1/2})\right\}\\
&=& \sigma v_{\theta}^{1/2} \E \left[\delta_{1-\alpha/2}\big\{\hat{\tau}/(\sigma v_{\tau}^{1/2}), \hat{\sigma}/\sigma \big\} - \delta_{\alpha/2} \big\{\hat{\tau}/(\sigma v_{\tau}^{1/2}), \hat{\sigma}/\sigma \big\} \right]\\
&=& \sigma \, v_{\theta}^{1/2} \int_{0}^{\infty}\int_{-\infty}^{\infty} \left\{\delta_{1-\alpha/2}(x, y) - \delta_{\alpha/2}(x, y) \right \}\phi(x-\gamma)\, f_{n-p}(y) \, dx \, dy.
\end{eqnarray*}

\subsection*{Proof of Theorem 3}

Henceforth, we make the dependence of $\delta_u(x,y)$ on $\rho$ explicit by using the notation $\delta_u(x,y, \rho)$ in place of
$\delta_u(x,y)$.
We first prove the following lemma.

\noindent
\textbf{Lemma 1.} \textit{$\delta_{1-\alpha/2}(-x, y, \rho) = -\delta_{\alpha/2}(x, y, \rho)$.}

\noindent \textit{Proof.}  From the definitions (\ref{eq:r2})--(\ref{eq:weight}), we have $r_1(\delta, -x, y)=-r_1(-\delta, x, y)$, $r_2(\delta, y)=-r_2(-\delta, y)$ and $w_1(x,y) = w_1(-x, y)$.  For any fixed $y > 0$, $\delta_{1-\alpha/2}(-x, y, \rho)$ is the solution in $\delta$ of
\begin{eqnarray*}
1-\alpha/2 &=& w_1(-x, y) \Phi\{r_1(\delta, -x, y)\} + \{1-w_1(-x, y)\} \Phi\{r_2(\delta, y)\}\\
&=& w_1(x, y)[1- \Phi\{r_1(-\delta, x, y)\}] + \{1-w_1(x, y)\} [1-\Phi\{r_2(-\delta, y)\}]\\
&=& 1 - \big[w_1(x, y)\Phi\{r_1(-\delta, x, y)\} + \{1-w_1(x, y)\}\Phi\{r_2(-\delta, y)\} \big],
\end{eqnarray*}
where the second line follows the fact that $\Phi(-x) = 1 -\Phi(x)$,
and hence of
\begin{eqnarray*}
\alpha/2 &=& w_1(x, y)\Phi\{r_1(-\delta, x, y)\} + \{1-w_1(x, y)\}\Phi\{r_2(-\delta, y)\}.
\end{eqnarray*}
It is therefore also \textit{minus} the solution in $\delta$ of
\begin{equation*}
\alpha/2
= w_1(x, y)\Phi\{r_1(\delta, x, y)\} + \{1-w_1(x, y)\}\Phi\{r_2(\delta, y)\}.
\end{equation*}
Since $\delta_{\alpha/2}(x,y, \rho)$ is the solution of this equation, we must have
$\delta_{1-\alpha/2}(-x, y, \rho)=-\delta_{\alpha/2}(x,y, \rho)$.

\bigskip \noindent
Proof: \textit{For the coverage}.

Let
\begin{eqnarray*}
C(\gamma,\rho) &=&
\int_{0}^{\infty} \int_{-\infty}^{\infty} \Bigg[ \Phi\Bigg\{\frac{\delta_{1-\alpha/2}(x,y,\rho) - \rho (x-\gamma)}{(1-\rho^2)^{1/2}} \Bigg\} \\
&& - \Phi\Bigg\{\frac{\delta_{\alpha/2}(x,y,\rho) - \rho (x-\gamma)}{(1-\rho^2)^{1/2}} \Bigg\}\Bigg]\phi(x-\gamma) \, f_{n-p}(y) \, dx \, dy.
\end{eqnarray*}
For each fixed $\rho$,
\begin{eqnarray*}
C(-\gamma,\rho) &=&
\int_{0}^{\infty} \int_{-\infty}^{\infty} \Bigg[ \Phi\Big\{\frac{\delta_{1-\alpha/2}(x,y,\rho) - \rho (x+\gamma)}{(1-\rho^2)^{1/2}} \Big\} \\
&& - \Phi\Big\{\frac{\delta_{\alpha/2}(x,y,\rho) - \rho (x+\gamma)}{(1-\rho^2)^{1/2}} \Big\}\Bigg]\phi(x+\gamma) \, f_{n-p}(y) \, dx \, dy\\
&=& \int_{0}^{\infty} \int_{-\infty}^{\infty} \Bigg[ \Phi\Big\{\frac{\delta_{1-\alpha/2}(-z,y,\rho) - \rho (-z+\gamma)}{(1-\rho^2)^{1/2}} \Big\} \\
&& - \Phi\Big\{\frac{\delta_{\alpha/2}(-z,y,\rho) - \rho (-z+\gamma)}{(1-\rho^2)^{1/2}} \Big\}\Bigg]\phi(-z+\gamma) \, f_{n-p}(y) \, dz \, dy\\
&=& \int_{0}^{\infty} \int_{-\infty}^{\infty} \Bigg[ \Phi\Big\{-\frac{\delta_{\alpha/2}(z,y,\rho) - \rho (z-\gamma)}{(1-\rho^2)^{1/2}} \Big\} \\
&& - \Phi\Big\{-\frac{\delta_{1-\alpha/2}(z,y,\rho) - \rho (z-\gamma)}{(1-\rho^2)^{1/2}} \Big\}\Bigg]\phi(z-\gamma) \, f_{n-p}(y) \, dz \, dy\\
&=& C(\gamma, \rho).
\end{eqnarray*}
The second line follows by changing the variable to $z=-x$, the third follows from the Lemma and the fact that the standard normal density is an even function, and the fourth follows from the fact that $\Phi(-x) = 1 -\Phi(x)$.

Using  \eqref{eq:h}, it is straightforward to show that $\delta_u(x,y,-\rho) = \delta_u(-x,y,\rho)$.
Thus, for each fixed $\gamma$,
\begin{eqnarray*}
C(\gamma,-\rho) &=&
\int_{0}^{\infty} \int_{-\infty}^{\infty} \Big[ \Phi\Big\{\frac{\delta_{1-\alpha/2}(-x,y, \rho) + \rho (x-\gamma)}{(1-\rho^2)^{1/2}} \Big\} \\
&& - \Phi\Big\{\frac{\delta_{\alpha/2}(-x,y, \rho) + \rho (x-\gamma)}{(1-\rho^2)^{1/2}} \Big\}\Big]\phi(x-\gamma) \, f_{n-p}(y) \, dx \, dy\\
&=&
\int_{0}^{\infty} \int_{-\infty}^{\infty} \Big[ \Phi\Big\{-\frac{\delta_{\alpha/2}(x,y, \rho) - \rho (x-\gamma)}{(1-\rho^2)^{1/2}} \Big\} \\
&& - \Phi\Big\{-\frac{\delta_{1-\alpha/2}(x,y, \rho) - \rho (x-\gamma)}{(1-\rho^2)^{1/2}} \Big\}\Big]\phi(x-\gamma) \, f_{n-p}(y) \, dx \, dy\\
&=& C(\gamma, \rho),
\end{eqnarray*}
where the second line follows from the Lemma and the third from the fact that $\Phi(-x) = 1 -\Phi(x)$.

\bigskip \noindent
Proof: \textit{For the expected length}.

The expected length and the scaled expected length are proportional to
\begin{eqnarray*}
L(\gamma, \rho) &=& \int_{0}^{\infty} \int_{-\infty}^{\infty} \{\delta_{1-\alpha/2}(x,y, \rho) - \delta_{\alpha/2}(x,y, \rho)\}\, \phi(x-\gamma) \, f_{n-p}(y) \, dx \, dy.
\end{eqnarray*}
For each fixed $\rho$,
\begin{eqnarray*}
L(-\gamma, \rho)
&=&  \int_{0}^{\infty} \int_{-\infty}^{\infty} \{\delta_{1-\alpha/2}(x,y, \rho) - \delta_{\alpha/2}(x,y, \rho)\} \, \phi(x+\gamma) \, f_{n-p}(y) \, dx \, dy\\
&=&  \int_{0}^{\infty} \int_{-\infty}^{\infty} \{\delta_{1-\alpha/2}(-z,y, \rho) - \delta_{\alpha/2}(-z,y, \rho)\}\, \phi(-z+\gamma) \, f_{n-p}(y) \, dz \, dy\\
&=& \int_{0}^{\infty} \int_{-\infty}^{\infty} \{-\delta_{\alpha/2}(x,y, \rho) + \delta_{1-\alpha/2}(z,y, \rho)\}\, \phi(z-\gamma) \, f_{n-p}(y) \, dz \, dy\\
&=& L(\gamma, \rho).
\end{eqnarray*}
The second line follows by changing the variable to $z=-x$ and the third follows from the Lemma and the fact that the standard normal density is an even function.

It follows from $\delta_u(x,y,-\rho) = \delta_u(-x,y,\rho)$ that
\begin{eqnarray*}
L(\gamma, -\rho) &=&  \int_{0}^{\infty} \int_{-\infty}^{\infty} \{\delta_{1-\alpha/2}(-x,y, \rho) - \delta_{\alpha/2}(-x,y, \rho)\}\, \phi(x-\gamma) \, f_{n-p}(y) \, dx \, dy\\
&=&  \int_{0}^{\infty} \int_{-\infty}^{\infty} \{-\delta_{\alpha/2}(x,y, \rho) + \delta_{1-\alpha/2}(x,y, \rho)\} \, \phi(x-\gamma) \, f_{n-p}(y) \, dx \, dy\\
&=& L(\gamma, \rho),
\end{eqnarray*}
where the second line follows from the Lemma.

\newpage

\clearpage

\newpage

\clearpage

\newpage

\clearpage

\newpage

\clearpage

\newpage

\clearpage

\newpage

\clearpage

\newpage

\end{document}